\documentclass[final,5p,times,twocolumn]{elsarticle}

\usepackage{amssymb}
\usepackage{amsmath}
\usepackage{amssymb}
\usepackage{graphicx}
\usepackage{longtable}

\journal{Chemical Physics Letters}

\begin{document}

\begin{frontmatter}

\title{Kramers turnover in class of thermodynamically open systems: Effect of interplay of nonlinearity and noises}

\author[besu]{Anindita Shit}

\author[besu]{Sudip Chattopadhyay\corref{cor1}}
\ead{sudip$_-$chattopadhyay@rediffmail.com}

\author[bi]{Suman Kumar Banik}
\ead{skbanik@bic.boseinst.ernet.in}

\author[katwa]{Jyotipratim Ray Chaudhuri\corref{cor1}}
\ead{jprc$_-$8@yahoo.com }

\cortext[cor1]{Corresponding author}

\address[besu]{Department of Chemistry, Bengal Engineering and Science University, 
Shibpur, Howrah 711103, India}

\address[bi]{Department of Chemistry, Bose Institute, 93/1 A P C
Road, Kolkata 700009, India}

\address[katwa]{Department of Physics, Katwa College, Katwa,
Burdwan-713130, India}

\begin{abstract}
A system-reservoir nonlinear coupling model has been proposed for
a situation where the reservoir is nonlinearly driven by an
external Gaussian stationary noise which exposes the system
particles to a nonequilibrium environment. Apart from the internal
thermal noise, the thermodynamically open system encounters two
other noises that are multiplicative in nature. Langevin equation
derived from the resulting composite system contains the essential
features of the interplay between these noise processes. Based on
the numerical simulation of the full model potential, we show that
one can recover the turnover features of the Kramers dynamics even
when the reservoir is modulated nonlinearly by an external noise.
\end{abstract}

\end{frontmatter}

\section{Introduction}

Inspired by a work of Christiansen \cite{chris}, where a chemical
reaction was considered as a diffusion problem,
Kramers \cite{kramers} introduced a Brownian motion model in a
one--dimensional (along the reaction coordinate) force field to
predict the existence of several kinetic regimes depending on the
magnitude of the friction (very low or energy diffusion regime,
and moderate to high or spatial diffusion regime). A clear
understanding of the pre-factor of the Kramers equation is useful
not only for completeness of the theory of escape rate, but also
for explaining various phenomena. Therefore, for the last few
decades, much effort has been put into extending the Kramers
model \cite{risken,ref1,ref8,ref82,ref8sc,ref8sc2,nitzan}. Many
authors have devised methods for obtaining escape rate in the
whole range of friction by extending the basic assumptions found
in the original Kramers work, known as the Kramers turnover
problem \cite{ref1,ref8,ref82,ref3,dykman}. Kramers has shown that
the rate constant is proportional to $\gamma$ (dissipation
constant) when $\gamma$ is low and proportional to $\gamma^{-1}$
when $\gamma$ is high. It can thus be expected that the value of
the rate constant reaches a maximum at an intermediate $\gamma$
and decreases to zero when $\gamma$ approaches either zero or
infinity. This dependence of the rate constant on friction is
known as Kramers turnover. However, Kramers could not derive a
uniform expression for the rate, valid for all values of the
friction coefficient. In fact, analytical solutions for Kramers
equation are only possible for very simple interaction
models \cite{risken,ref1,ref82}. The systematic solution of the
Kramers turnover problem for the thermodynamically closed system
was given by Pollak--Grabert--H\"{a}nggi(PGH) \cite{PGH} (one of
the foremost studies about turnover) who generalized the Kramers
model to an arbitrary time-dependent friction and demonstrated
that the turnover formula due to Mel'nikov and
Meshkov \cite{ref8,melnikov} can be obtained without any {\it ad
hoc} bridging. Later, the PGH theory was generalized to many
dimensions \cite{ref15,hershkovitz}.

Activated rate processes in one-dimensional surface diffusion have
been studied by Pollak {\em et al.} \cite{ref15,pollak}.
Hershkovitz and Pollak \cite{hershkovitz} studied the length
dependence of the classical activated transfer rate across a
bridge and found that the Kramers turnover theory in the rate
suffices for understanding the bridge length and friction
dependence of the rate. Rips and Pollak \cite{ripspollak} extended
the PGH method in the context of quantum Kramers turnover problem.
Segal {\em et al.} \cite{segal} have provided the first analysis of
the transition from the tunnelling to the thermally activated
regime in a variant of the quantum Kramers problem as a function
of the barrier length. Vega {\em et al.} \cite{vega} have studied
the Kramers turnover theory in activated atom–-surface diffusion
using mean first passage time. Shepherd and
Hernandez \cite{SHjcp115,SHjcp117,SHjpc106} have exploited the mean
first passage time (MFPT) based rate formula to analyze the
interplay between Kramers turnover and resonant activation for the
escape rates on stochastic bistable and aperiodic potentials. The
development in Ref.~\cite{SHjpc106} is particularly
interesting as it investigates the low friction regime (the most
difficult part of Kramers turnover theory to illustrate) in
conjunction with stochastic aperiodic potentials. Recently,
Kramers turnover has also been realized during the investigation
of the forward and backward reaction rates of the
LiNC$\rightleftharpoons $LiCN isomerization reaction in a bath of
argon atoms at various densities using molecular dynamics
simulations due to Garc\'{i}a-M\"{u}ller {\it et al.} \cite{prl08}.
Their work provides clear evidence for the increase in rates with
microscopic friction in the energy-diffusion regime in chemical
system.

The last few decades have observed a crescendo of research
activity in the field of nonequilibrium statistical mechanics
using the system-reservoir (SR) model
\cite{lindenbergwest,srzwanzig}. In the overwhelming majority of
situations, the interaction between the system and the reservoir
has been considered to be linear in bath co-ordinates as well
system co-ordinates. This in turn relates the additive noise of
the thermal bath with linear dissipation of the system through
fluctuation--dissipation relation (FDR). On the other hand, if the
SR coupling is nonlinear in system coordinate, the corresponding
Hamiltonian gives rise to a Langevin equation with state-dependent
dissipation and internal multiplicative thermal noise. As the
total SR combination is thermodynamically closed, the energy
balance condition again is reflected through FDR \cite{vulpiani}.
If density of bath modes is such that the associated noise is
stationary and Gaussian, one can numerically solve the associated
Langevin equation for barrier crossing dynamics to observe the
turnover phenomena. It should be recognized that it is very hard
to obtain a simple expression for escape rate, even for a white
noise process, when the dissipation is state dependent and the
noise process is multiplicative in nature. Consequently,
PGH \cite{PGH}-type analysis for turnover problem is very hard to
achieve in such cases \cite{prl08}. At this point, one might wonder
if the turnover phenomena can be observed when the SR combination
is thermodynamically open and hence there is no energy-balance
relation like FDR.

Among many other situations, the SR combination will be
thermodynamically open if one drives the system externally
(keeping the reservoir in thermal equilibrium). On the other hand,
in spite of direct driving, one may expose the reservoir to an
external modulation. A number of different situations depicting
the modulation of the bath by an external noise may be physically
relevant \cite{dsrjrc,jrcbath,jrcbath2,jrcsb}. Whether the system
or the reservoir is driven by an external noise, there is an
additional mechanism to inject energy into the system and clearly,
there is no FDR in such a situation. Inspection of any such
situation may be relevant to examine the turnover phenomena in the
rate. In what follows, we address the later situation where the
reservoir is modulated externally by a random force to make the SR
combination thermodynamically open. It is our aim here to search
for the signature of Kramers turnover in the rate, emerging from
the nonlinear driving of the bath by an external noise. The effect
of nonlinear modulation of the reservoir by an external agency is
considered indispensable in explaining the phenomena of activation
of a quasibound species (reactants surrounded by the solvent
molecules) above its trapping potential in the presence of high
intensity light sources. For example, one may consider an
isomerization reaction (A $\rightleftharpoons$ B) in a
photochemically active solvent in the presence of an external
light source with high intensity.

\section{Methodology}\label{theory}

To start with, we consider a classical particle of unit mass being
coupled to a heat bath consisting of ${\mbox N}$-mass weighted
harmonic oscillators, characterized by the frequency set
$\{\omega_j\}$ (i.e. the bath degrees of freedom are described by
an ensemble of oscillators). In addition to that, the heat bath is
nonlinearly driven by an external noise identified as
$\epsilon(t)$. The Hamiltonian for the composite system is
\begin{eqnarray} \label{eq1}
{\mbox H} = {\mbox H}_{\rm S} + {\mbox H}_{\rm B} + {\mbox H}_{\rm
SB}+ {\mbox H}_{\rm int}
\end{eqnarray}
where the Hamiltonian of the system is expressed as: ${\mbox
H}_{\rm S}=\left(p^2/2\right)+V(q)$ with $V(q)$ being the
potential energy function and $p$ and $q$ being respectively the
coordinate and the momentum of the system particle. ${\mbox
H}_{\rm B} + {\mbox H}_{\rm SB} = \sum_{j=1}^N \left [ \frac
{p_j^2}{2} + \frac{1}{2}\omega_j^2\left\{x_j-c_j f\left(
q\right)\right\}^2\right]$ where $\{x_j,p_j\}$ are the variables
for the $j$-th bath oscillator. The system-heat bath interaction
is given by the coupling term $c_j\omega_j f(q)$ where $c_j$ is
the coupling strength and $f(q)$ is some well-behaved function of
the system coordinate $q$ only. Through the insertion of the term
$f(q)$, we have considered the SR interaction to be, in general,
nonlinear. For bilinear system-bath coupling, $f(q)$ would have
been taken as some linear function of $q$. The interaction between
the heat bath and external noise $\epsilon(t)$ is taken as ${\mbox
H}_{\rm int}=\sum_{j=1}^N \kappa_j g\left({x}_j\right) \epsilon
(t)$ where $\kappa_j$ denotes the strength of the interaction and
$g(x_j)$ is an arbitrary analytic function of the bath variable
$x_j$. This type of interaction makes the bath variables
explicitly time dependent. A large class of phenomenologically
modelled stochastic differential equations may be obtained from a
microscopic Hamiltonian for a particular choice of coupling
function $g(x_j)$ and have already been used for microscopic
realization of Kubo-type oscillator and correlated noise processes
\cite{kubo,kubo2}. If one chooses, for example,
$g(x_j)=\frac{1}{2}x_j^2$, the spring constants of the bath
oscillators become fluctuating. In what follows, we choose
$g(x_j)=x_j+\frac{1}{2}x_j^2$; a linear-linear(LL) and a
square-linear(SL) coupling between the noise and bath variables.
Recently, the effect of such LL coupling and SL coupling between
the system and the reservoir have been studied by Tanimura and
coworkers\cite{tanimura} in the context of spectroscopic studies.
The notation of the hierarchical representation of the underlying
Hamiltonian system in Eq.(\ref{eq1}) and its subsequent analytic
representations in extended Langevin equations has been discussed
by Popov and Hernandez \cite{SHjcp126}.

In what follows the external noise $\epsilon(t)$ is taken to be
stationary, Gaussian with statistical properties $\langle \epsilon (t) \rangle_e = 0$
\begin{eqnarray}\label{eq2}
\langle \epsilon (t) \epsilon (t^\prime) \rangle_e =
\psi(t-t^\prime) =
\frac{D_\epsilon}{\tau_\epsilon}\exp\left(-\frac{|t-t^\prime|}{\tau_\epsilon}\right) ,
\end{eqnarray}

\noindent where $D_\epsilon$ is the strength of the noise and
$\tau_\epsilon$ is its correlation time. For
$\tau_\epsilon\rightarrow 0$, $\epsilon(t)$ becomes
$\delta$-correlated with statistical property: $\langle \epsilon
(t) \epsilon (t^\prime) \rangle_\epsilon = 2
D_\epsilon\delta(t-t^\prime)$. In Eq.(\ref{eq2}),
$\langle\cdots\rangle_e$ implies averaging over the external noise
processes. In this context, we want to mention that the presence
of noise and nonlinearity are unavoidable in general physical
systems, So, one must take into account the interplay between
these two factors on the dynamics of the system.

From Eq.~(\ref{eq1}), we have the dynamical equations for the
system and the bath variables as
\begin{eqnarray}
\ddot{q}(t)=-V^\prime(q(t))+f^\prime(q(t))\sum_{j} c_j
\omega_j^2 \left\{ x_j(t)-c_j f(q(t))\right\} \label{eq3}, \\
\ddot{x}_j(t)+\left\{\omega_j^2+\kappa_j\epsilon(t)\right\}x_j(t)=-\kappa_j\epsilon(t)+
c_j \omega_j^2f(q(t))\label{eq4}.
\end{eqnarray}

\noindent
To solve Eq.(\ref{eq4}) for $x_j(t)$, we assume a solution of the form
\begin{eqnarray}\label{eq5}
x_j(t)=x_j^0(t)+\kappa_jx_j^1(t), 
\end{eqnarray}

\noindent
where $x_j^0(t)$ is the solution of the unperturbed equation of
motion (EOM)
\begin{eqnarray}\label{eq6}
\ddot{x}_j^0(t)+\omega_j^2x_j^0(t) &= c_j \omega_j^2 f(q(t)) .
\end{eqnarray}

\noindent We now consider that at $t=0$, the heat bath is in thermal
equilibrium in the presence of the system but in the absence of
the external noise $\epsilon(t)$. Subsequently, at $t=0_+$, the
external noise agency is switched on and the heat bath is
modulated by $\epsilon(t)$. Then $x_j^1(t)$ must satisfy the
equation
\begin{eqnarray}\label{eq7}
\ddot{x}_j^1(t)+\omega_j^2x_j^1(t)
=-\epsilon(t)-\epsilon(t)x_j^0(t) .
\end{eqnarray}

\noindent with the initial conditions $x_j^1(0)=p_j^1(0)=0$. Now, the
solution of Eq.(\ref{eq7}) is given by
\begin{eqnarray}\label{eq8}
x_j^1(t) &=& -\frac{1}{\omega_j}\int_0^tdt^\prime \sin
\omega_j(t-t^\prime)\epsilon(t^\prime)\nonumber \\
&&-\frac{1}{\omega_j}\int_0^tdt^\prime
\sin \omega_j(t-t^\prime)x_j^0(t^\prime)\epsilon(t^\prime) .
\end{eqnarray}

\noindent The formal solution of Eq.(\ref{eq6}) is given by
\begin{eqnarray} \label{eq9}
x_j^0(t)&=&x_j^0(0)\cos\omega_j
(t)+\frac{p_j^0(0)}{\omega_j}\sin\omega_j
(t) \nonumber \\
&& +c_j\omega_j\int_0^tdt^\prime \sin
\omega_j(t-t^\prime)f(q(t^\prime)),
\end{eqnarray}

\noindent where $x_j^0(0)$ and $p_j^0(0)$ are respectively the initial
position and momentum of the $j$-th bath oscillator. Now, using
this solution in Eq.(\ref{eq8}), we have (after an integration by
parts) the EOM for bath variables $x_j(t)$ [from Eq.(\ref{eq5})]
as
\begin{eqnarray} \label{eq10}
x_j(t)-c_jf(q(t))&=&\left\{x_j^0(0)-c_jf(q(0))\right\}\cos\omega_j
(t)\nonumber \\
&&+\frac{p_j^0(0)}{\omega_j}\sin\omega_j t \nonumber \\
&& -c_j\int_0^t dt^\prime\cos\omega_j
(t-t^\prime)f^\prime(q(t^\prime))\dot{q}(t^\prime)\nonumber\\
&&-\frac{\kappa_j}{\omega_j}\int_0^t dt^\prime\sin
\omega_j(t-t^\prime)\epsilon(t^\prime)\nonumber \\
&&-\frac{\kappa_j}{\omega_j}\int_0^t
dt^\prime\sin
\omega_j(t-t^\prime)x_j^0(t^\prime)\epsilon(t^\prime). \nonumber \\
\end{eqnarray}

\noindent
Using the above solution in Eq.(\ref{eq3}), we obtain the EOM for the system variables as
\begin{eqnarray}\label{eq11}
\ddot{q}(t)&=&-V^\prime(q(t))-f^\prime(q(t))\int_0^tdt^\prime
\gamma(t-t^\prime)f^\prime(q(t^\prime))\dot{q}(t^\prime)\nonumber \\
&&+f^\prime(q(t))
F(t)+f^\prime(q(t))\pi(t)-f^\prime(q(t)) \nonumber \\
&& \times \int_0^tdt^\prime\left\{\sum_j
c_j\kappa_j\omega_j\sin\omega_j(t-t^\prime)x_j^0(t^\prime)\right\}\epsilon(t^\prime),
\nonumber \\
\end{eqnarray}

\noindent where the damping kernel is given by $\gamma(t)=\sum_j
c_j\omega_j^2\cos\omega_jt$. F(t) is the internal thermal noise
generated through the coupling between the system and the heat
bath and is given by
\begin{eqnarray}
F(t)&=& \sum_j c_j \omega_j^2 \left[\left\{ x_j^0(0)- c_j
f(q(0))\right\} \cos \omega_jt \right. \nonumber \\
&& \left.+ \frac{p_j^0(0)}{\omega_j} \sin
\omega_j t\right],
\end{eqnarray}

\noindent and
\begin{eqnarray}\label{scjrc2}
\pi(t)=\int_0^tdt^\prime\varphi(t-t^\prime)\epsilon(t^\prime) ,
\end{eqnarray}

\noindent is a dressed noise that depends on the external noise
$\epsilon(t)$ and
\begin{eqnarray}\label{scjrc3}
\varphi(t)=\sum_jc_j\kappa_j\omega_j\sin\omega_j t.
\end{eqnarray}

\noindent Clearly, the system does not encounter the external noise
$\epsilon(t)$ directly, rather, the driving of the bath by the
external noise $\epsilon(t)$ results in a dressed noise. The form
of Eq.(\ref{eq11}) therefore suggests that the system is driven by
two forcing functions ${\mbox F(t)}$ and $\pi(t)$. ${\mbox F(t)}$
depends on the initial conditions of the bath oscillators for a
fixed choice of the initial condition of the system degrees of
freedom. To define the statistical properties of ${\mbox F(t)}$,
we assume that the initial distribution is the one in which the bath
is equilibrated at ${\mbox t}=0$ in the presence of the system but
in the absence of the external noise agency. Let us now digress a
little bit about $\pi(t)$. The statistical properties of $\pi(t)$
are determined by the normal-mode density of the bath frequencies,
the coupling of the system with the bath, the coupling of the bath
with the external noise, and the external noise itself. Equation
(\ref{scjrc2}) is reminiscent of the familiar linear relation
between the polarization and the external field, where $\pi$ and
$\epsilon$ play the role of the former and the latter,
respectively. The function $\varphi(t)$, thus may be taken as the
response function of the bath. The very structure of $\pi(t)$
suggests that this forcing function, although originating from an
external force, is different from a direct driving force acting on
the system. The distinction lies at the very nature of the bath
characteristics (rather than system characteristics) as reflected
in the relations Eqs.~(\ref{scjrc2}) and ~(\ref{scjrc3}). At this
point, we note that the forcing term $F(t)$ is deterministic. It
ceases to be deterministic if it is not possible to specify all
the $x_j^0(0)$'s and $p_j^0(0)$'s, i.e., the initial conditions of
all the bath variables, exactly. The standard procedure to
overcome this difficulty is to consider a distribution of
$x_j^0(0)$ and $p_j^0(0)$ to specify the statistical properties of
the bath-dependent forcing term $F(t)$. The distribution of the
bath oscillators is assumed to be a canonical distribution of the
Gaussian form
\begin{eqnarray} \label{eq12}
\rho_{\rm eq}^{\rm bath}(0)&=&N \exp\left[-\frac{1}{k_BT}\left\{
\sum_{j}\left(\frac{{p_j^0}^2(0)}{2} \right. \right. \right. \nonumber \\
&& \left. \left. \left. +\frac{1}{2}\omega_j^2(x_j^0(0)-c_jf(q(0)))^2\right)\right\}\right] ,
\end{eqnarray}

\noindent where $N$ is the normalization constant. This choice of the
distribution function of the bath variables makes the initial
noise Gaussian. It is now easy to verify the statistical
properties of $F(t)$ as $\langle F(t)\rangle=0$ and $\langle
F(t)F(t^\prime)\rangle=k_BT \gamma(t-t^\prime)$ where $k_B$ is the
Boltzmann constant and $T$ is the equilibrium temperature. Here,
$\langle\cdots\rangle$ implies the average over the initial
distribution given in Eq.(\ref{eq12}). The second relation is the
FDR \cite{ref3} which ensures that the
bath was in thermal equilibrium at $t=0$, in presence of the
system. To proceed further, we consider the last term of
Eq.(\ref{eq11}) as
\begin{eqnarray} \label{eq12a}
\Gamma(t)=f^\prime(q(t))\int_0^tdt^\prime\sum_j
c_j\kappa_j\omega_j\sin\omega_j(t-t^\prime)\epsilon(t^\prime)x_j^0(t^\prime) .
\end{eqnarray}

\noindent We now put the expression for $x_j^0(t^\prime)$ from
Eq.(\ref{eq9}). The solution Eq.(\ref{eq9}), consists of two
parts, the homogeneous solution of Eq.(\ref{eq6}) which is the
free evolution of bath variables is the fast part. The second one
is the solution of the corresponding inhomogeneous equation which
gives the the forced oscillation expressed as
$c_j\omega_j\int_0^tdt^\prime\sin\omega_j(t-t^\prime)f(q(t^\prime))$.
As the fast part dies out quickly for damped driven oscillator, we
pick the particular solution of Eq.(\ref{eq9}) only for $x_j^0(t)$
and consequently, Eq.(\ref{eq11}) becomes
\begin{eqnarray}\label{eq13}
\ddot{q}(t)&=&-V^\prime(q(t))-f^\prime(q(t))\int_0^tdt^\prime
\gamma(t-t^\prime)f^\prime(q(t^\prime))\dot{q}(t^\prime) \nonumber \\
&&+f^\prime(q(t))F(t)+f^\prime(q(t))
\pi(t)\nonumber\\
&& -f^\prime(q(t))\int_0^tdt^\prime\epsilon(t^\prime)\int_0^{t^\prime}
dt^{\prime\prime}f(q(t^{\prime\prime})) \nonumber \\
&& \times \left\{\sum_j
c_j^2\kappa_j\omega_j^2\sin\omega_j(t-t^\prime)\sin\omega_j(t^\prime-t^{\prime\prime})\right\}.
\end{eqnarray}

To identify Eq.(\ref{eq13}) as a generalized Langevin equation, we
must impose some conditions on the coupling coefficients $c_j$ and
$\kappa_j$, on the bath frequencies $\omega_j$ and on the number
${\mbox N}$ of the bath oscillators that will ensure that
$\gamma(t)$ is indeed dissipative and the last term in
Eq.(\ref{eq13}) is finite for $N\rightarrow\infty$. A sufficient
condition for $\gamma(t)$ to be dissipative is that it is
positive--definite and decreases monotonically with time. These
conditions are achieved if $N\rightarrow \infty$ and if $c_j
\omega_j^2$ and $\omega_j$ are sufficiently smooth functions of
$j$\cite{jmp51}. As ${\mbox N} \rightarrow \infty$, one replaces
the sum by an integral over $\omega$ weighted by a density of
states $\rho(\omega)$. Thus, to obtain a finite result in the
continuum limit, the coupling function $c_j=c(\omega)$ and
$\kappa_j=\kappa(\omega)$ are chosen as
$c(\omega)=\frac{c_0}{\omega\sqrt{\tau_c}}$ and
$\kappa(\omega)=\kappa_0$ where $c_0$ and $\kappa_0$ are constants
and $\tau_c$ is the correlation time of the heat bath. The choice
$\kappa(\omega)=\kappa_0$ is the simplest one where we assume that
every bath mode is excited with the same intensity. This simple
choice makes the relevant term finite for $N\rightarrow\infty$.
Consequently, $\gamma(t)$ becomes
\begin{eqnarray}
\gamma(t)=\left(\frac{c_0^2}{\tau_c}\right)d\omega
\rho(\omega)\cos\omega t,
\end{eqnarray}

\noindent where $1/\tau_c$ may be characterized as the cut-off frequency
of bath oscillators. The density of modes of $\rho(\omega)$ of the
heat bath is assumed to be Lorentzian,
\begin{eqnarray}
\rho(\omega)=\left(\frac{2}{\pi}\right)\left[\frac{\omega^2}{\tau_c^{-2}+\omega^2}\right].
\end{eqnarray}

\noindent This type of choice of $\rho(\omega)$ may be encountered in many
situations in chemical physics and condensed matter
physics\cite{sr2,sr4,sr5,coffey} and resembles broadly, in behavior,
the hydrodynamic modes in certain macroscopic
systems\cite{jcp16}. With these forms of $\rho(\omega)$,
$c(\omega)$ and $\kappa(\omega)$, we have the expression for
$\gamma(t)$ as
$\gamma(t)=\frac{c_0^2}{\tau_c}\exp\left(-\frac{t}{\tau_c}\right)$
which reduces to $\gamma(t)=2c_0^2\delta(t)=2\gamma\delta(t)$ for
$\tau_c\rightarrow 0$ where $\gamma=c_0^2$ and is a Markovian
dissipation constant and consequently, one obtains
$\delta$-correlated internal noise processes. With these forms of
density of modes $\rho(\omega)$ and coupling functions,
$c(\omega)$ and $\kappa(\omega)$, the response function
$\varphi(t)$ can be written in the continuum limit as
\begin{eqnarray} \label{eq13a}
\varphi(t)&=&\int_0^\infty d\omega \rho
(\omega)c(\omega)\kappa(\omega)\omega\sin \omega t \nonumber \\
&=&\frac{2}{\pi}c_0\kappa_0\frac{1}{\tau_c}\int_0^\infty d\omega
\omega\frac{\sin \omega t}{\tau_c^{-2}+\omega^2} \nonumber \\
&=&\frac{c_0\kappa_0}{\tau_c}\exp\left(\frac{-t}{\tau_c}\right).
\end{eqnarray}

\noindent Clearly, for $\tau_c\rightarrow 0$, $\varphi(t)$ reduces to
$\varphi(t)=2c_0\kappa_0\delta(t)$. Now, using the standard
trigonometric identity, the last term in Eq.(\ref{eq13}) can be
written as
\begin{eqnarray} \label{eq14}
\Delta(t)&=&f^\prime(q(t))\left[\frac{1}{2}\int_0^t dt^\prime
\epsilon (t^\prime)\int_0^{t^\prime}
dt^{\prime\prime}f[q(t^{\prime\prime})] \right. \nonumber \\
&& \times \sum_jc_j^2
\kappa_j\omega_j^2\cos
\omega_j(t-2t^\prime+t^{\prime\prime}) \nonumber\\
&& -\frac{1}{2}\int_0^t dt^\prime \epsilon
(t^\prime)\int_0^{t^\prime} dt^{\prime\prime}
f[q(t^{\prime\prime})] \nonumber \\
&& \left. \times \sum_jc_j^2 \kappa_j\omega_j^2\cos
\omega_j(t-t^\prime)\right].
\end{eqnarray}

\noindent Now, using the assumed expressions for the coupling functions
$c(\omega)$ and $\kappa(\omega)$ and the density of modes
$\rho(\omega)$, one easily observes that the two sums in
Eq.(\ref{eq14}) may be approximated as a $\delta$-function,
\begin{eqnarray} \label{eq15}
\sum_jc_j^2 \kappa_j\omega_j^2\cos
\omega_j(t+t^{\prime\prime}-2t^\prime)&=&\int d\omega \rho(\omega)
\left\{c(\omega)\right\}^2\kappa(\omega) \nonumber \\
&& \times \omega^2\cos\omega(t+t^{\prime\prime}-2t^\prime)
\nonumber\\&=&2c_0^2 \kappa_0\delta(t+t^{\prime\prime}-2t^\prime).
\end{eqnarray}

\noindent Similarly,
\begin{eqnarray} \label{eq16}
\sum_jc_j^2 \kappa_j\omega_j^2\cos
\omega_j(t-t^{\prime\prime})=2c_0^2
\kappa_0\delta(t-t^{\prime\prime}) .
\end{eqnarray}

\noindent Thus, in the continuum limit, the expression for $\Delta(t)$ reduces to
\begin{eqnarray} \label{eq17}
\Delta(t)&=&c_0^2 \kappa_0f^\prime(q(t))\left[\int_0^t dt^\prime
\epsilon (t^\prime)\int_0^{t^\prime}
dt^{\prime\prime}f[q(t^{\prime\prime})] \right. \nonumber \\
&& \times \delta(t+t^{\prime\prime}-2t^\prime) \nonumber \\
&& \left. -
\int_0^t dt^\prime \epsilon (t^\prime)\int_0^{t^\prime}
dt^{\prime\prime}f[q(t^{\prime\prime})]\delta(t-t^{\prime\prime})\right].
\end{eqnarray}

\noindent With the property of $\delta$-function, the first double integral in 
Eq.(\ref{eq17}) may be written as
\begin{eqnarray} \label{eq18}
&& \int_0^t dt^\prime \epsilon (t^\prime)\int_0^{t^\prime}
dt^{\prime\prime}f[q(t^{\prime\prime})]\delta(t+t^{\prime\prime}-2t^\prime) \nonumber \\
&& =\frac{1}{2}\int_0^tdy\epsilon\left(\frac{y+t}{2}\right)f(q(y)) .
\end{eqnarray}

\noindent As the system variable evolves much slowly in comparison to the
external noise $\epsilon(t)$, the right hand side of
Eq.(\ref{eq18}) may be approximated as
$\frac{1}{2}\left[\int_0^tdy\epsilon\left(\frac{y+t}{2}\right)\right]f(q(0))$.
For large $t$, we note that as
\begin{eqnarray}
\lim_{t\to\infty}\frac{1}{t} \int_0^t dt^\prime \epsilon
(t^\prime)=\langle\epsilon(t)\rangle_e=0,
\end{eqnarray}
the first term in the expression of $\Delta (t)$ vanishes.

To perform the second integration; $\int_0^t
dt^{\prime}\epsilon(t^{\prime})\int_0^{t^\prime}
dt^{\prime\prime}f[q(t^{\prime\prime})]\delta(t-t^{\prime})$, we
consider the region of integration, shown as the shaded triangle
in Figure~1. From the property of $\delta$-function, one observes
that the above integral will contribute only when
$t^{\prime\prime}=t$ but the inequality $0\leq
t^{\prime\prime}\leq t^{\prime}\leq t$ demands that at the same
time, $t^{\prime}$ should be equal to $t$. Thus, the contribution
from the integral come out only at point $P$ and the value of this
contribution is $f(q(t))\epsilon(t)$. Using all the above
facts, we obtain from Eq.(\ref{eq13}) the EOM for
system variables, in the limit $\tau_c\rightarrow 0$, as
\begin{eqnarray}\label{eq19}
\ddot{q}(t)&=&-V^\prime(q(t))- \gamma
[f^\prime(q)]^2p+f^\prime(q(t))F(t) \nonumber \\
&& +f^\prime(q(t))\pi(t)+\gamma\kappa_0f(q)f^\prime(q)\epsilon(t) .
\end{eqnarray}

\noindent This equation can be used to explore the distinctive aspects of
the reservoir (driven nonlinearly by an external noise) modulated
dynamics of the system in contrast to direct driving of the system
by the external noise. This will help us to elucidate the special
role of the reservoir response function in controlling the escape
of a Brownian particle from the metastable state.


\begin{figure}[!t]
\begin{center}
\includegraphics[width=1.0\columnwidth,angle=0]{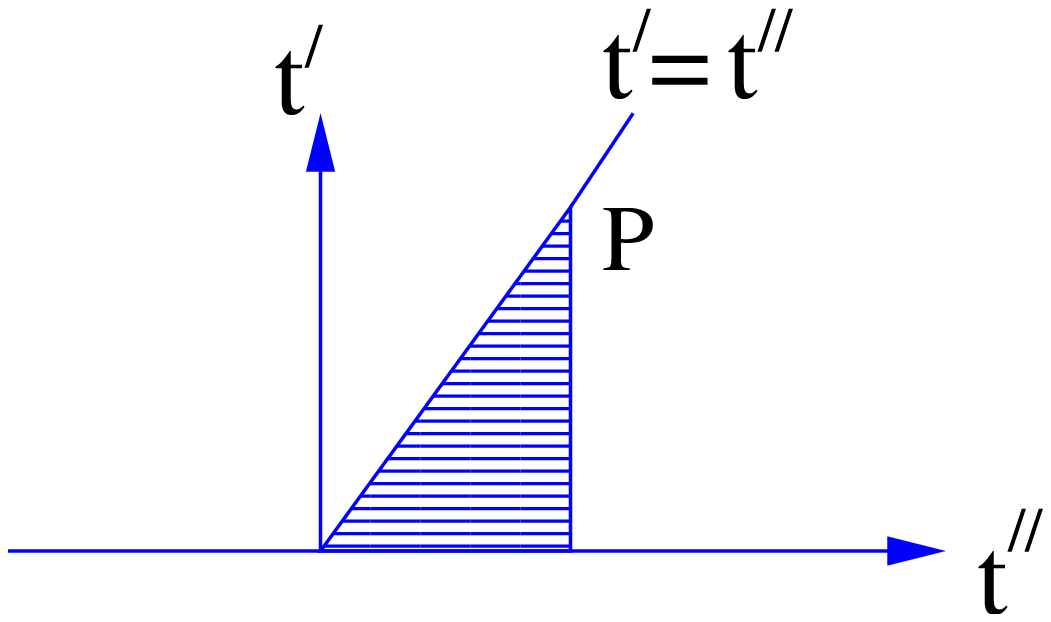}
\end{center}
\caption{Domain of integration of $\int_0^t
dt^{\prime}\epsilon(t^{\prime})\int_0^{t^\prime}
dt^{\prime\prime}f[q(t^{\prime\prime})]\delta(t-t^{\prime})$ in
Eq. (\ref{eq17}).
}
\label{fig1}
\end{figure}

\section {Results and Discussion: Kramers turnover}\label{theory2}

Before examining the noise induced transport, it is instructive
here to have a close look at the above Langevin equation, where
three noise processes appear and all these noise processes appear
multiplicatively. $F(t)$ is the internal thermal noise for which
FDR exists. $\pi(t)$ is the dressed noise and $\epsilon(t)$ is the
external noise. Instead of nonlinear SR coupling,
if one considers bilinear coupling, i.e., $f(q)=q$, the above
equation Eq.(\ref{eq19}) reads as
\begin{eqnarray}\label{eq20}
\ddot{q}(t)=-V^\prime(q(t))- \gamma p+F(t)+\pi(t)+\gamma\kappa_0
q\epsilon(t),
\end{eqnarray}

\noindent which indicates that both the thermal noise and dressed noise
appear additively but the last noise containing term appears
multiplicatively. The effect of interference of colored additive
and multiplicative white noises on escape rate has also been
explored using this type of equation\cite{bag}. Let us now discuss
a little bit on the origin of the noises appeared in
Eq.(\ref{eq20}). $F(t)$, the usual thermal noise appears due to
the system-bath interaction. The driving of the reservoir by
external noise yields the last two terms in Eq.(\ref{eq20}). If we
choose the bath-noise coupling function $g(x_j)$ to be linear in
bath variable, one will encounter the $\pi(t)$ noise in
Eq.(\ref{eq20}) only and the last term will disappear. On the
other hand, if $g(x_j)$ be quadratic, i.e., $g(x_j)=(1/2)x_j^2$,
$\pi(t)$ term disappears and the last term plays its role in the
dynamics. Here, it is interesting to note that the multiplicative
nature of the last noise process stems from the nonlinear driving
of the bath but not from the nonlinearity of system-reservoir
coupling function, which is the case for the other two noises.
Here, we enunciate a system without proof that if the bath-noise
coupling function be $g(x_j)=ax_j+bx_j^2+cx_j^3+...$, then for
linear system reservoir coupling, the resulting Langevin equation
will read as
\begin{eqnarray}\label{eq21}
\ddot{q}(t)=-V^\prime(q(t))- \gamma p+F(t)+\pi(t)+B
q\epsilon(t)+Cq^2\epsilon(t).
\end{eqnarray}

At this point, it is instructive to consider the statistical property of the
dressed noise $\pi(t)$ which can be easily verified as $\langle \pi (t) \rangle_e=0$
and
\begin{eqnarray}\label{eq22}
\langle \pi (t) \pi (t^\prime) \rangle_e &=& \int_0^t
dt^{\prime\prime}\int_0^{t^\prime}
dt^{\prime\prime\prime}\varphi(t-t^{\prime\prime})
\varphi(t^\prime-t^{\prime\prime\prime}) \nonumber \\
&& \times \psi(t^{\prime\prime}-t^{\prime\prime\prime}).
\end{eqnarray}

\noindent If we assume that the external noise $\epsilon(t)$ is
$\delta$-correlated, i.e., $\langle \epsilon(t) \epsilon
(t^\prime) \rangle= 2 D_e \delta(t-t^\prime)$, then in the limit
$\tau_c\rightarrow0$, the correlation function of $\pi (t)$
becomes $\langle \pi (t) \pi (t^\prime) \rangle_e=2
\gamma\kappa_0^2D_e\delta(t-t^\prime) $. In passing, we observe
that the system encounters an effective Gaussian additive noise
$\xi(t)$ $[=F(t)+\pi(t)]$ and another noise which appears
multiplicatively. The noises $\xi(t)$ and $\epsilon(t)$ are not
statistically independent, their correlation may be expressed as
$\langle \xi (t) \epsilon (t^\prime) \rangle=\langle \epsilon (t)
\xi (t^\prime) \rangle=\beta(t-t^\prime) $ which one may calculate
for a particular $\psi(t)$. Thus, the two mutually correlated
noises appear in the dynamical equation of the open system. The
appearance of cross-correlated noises has already been encountered
while explaining various physical
phenomena\cite{jrcsb,jrcpccp,jrcpccp2,newjp3}. Now, in terms of an
auxiliary function G(q) and a Gaussian stationary noise $R(t)$,
the Langevin equation Eq.(\ref{eq20}) can be written as
\begin{eqnarray}\label{eq23}
\ddot{q}(t)=-V^\prime(q(t))- \gamma p+G(q)R(t),
\end{eqnarray}

\noindent with
\begin{eqnarray}\label{eq24}
\langle\langle R (t) \rangle_e = 0 \;
\langle\langle R (t) R (t^\prime) \rangle\rangle 
=2 \delta (t-t^\prime), 
\end{eqnarray}

\noindent where $\langle\langle\cdots\rangle\rangle$ implies average over
the noise process $R(t)$(this averaging over $R(t)$ consists of
two independent averaging, one over thermal noise $F(T)$ and
another over external noise $\epsilon(t)$). In Eq.(\ref{eq23}),
\begin{eqnarray}\label{eq25}
G(q)=\left[(\gamma k_BT+\gamma \kappa_0^2
D_e)+D_e\gamma^2\kappa_0^2q^2+2\gamma\kappa_0D_e
q\right]^{1/2} .
\end{eqnarray}

\noindent
Clearly, Eq.(\ref{eq22}) along with Eq.(\ref{eq23}), is not the
FDR but serves as the thermodynamic consistency relation.

We now proceed to examine the noise induced transport. To do this,
we numerically solve the Langevin equation, Eq.(\ref{eq20})
[considering only quadratic bath-noise coupling, $g(x_j)=bx_j^2$],
by the method developed by Sancho {\it et al.} for multiplicative
noise and routinely calculate the MFPT\cite{mfpt}, the inverse of
which gives the escape rate from the metastable potential well. In
our numerical implementation, we consider a double well potential
of the form:
\begin{eqnarray*}
V(q)=-\frac{\rm A}{2}q^2+\frac{\rm B}{4}q^4 ,\;\;\; -\infty < q
<+\infty
\end{eqnarray*}


\begin{figure}[!t]
\begin{center}
\includegraphics[width=0.75\columnwidth,angle=-90]{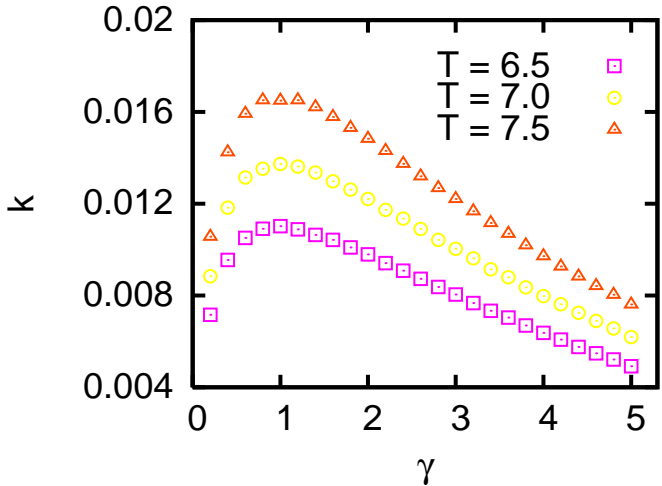}
\end{center}
\caption{Change of the transition rate, $k$ (sec$^{-1}$) with
the dissipation constant, $\gamma$ (sec$^{-1}$) for various
temperatures in conjunction with k$_B$=1, $A$=3.0, $B$ =0.1
$\kappa_0$=0.05, and $D_\epsilon$ =5.0.
}
\label{fig2}
\end{figure}

In Figure~2, we have plotted the rate k, obtained from Langevin
simulation using the concept of mean first passage time, as a
function of dissipation constant($\gamma$) for various
temperatures. For small $\gamma$, we observe that the rate
increases with increase in $\gamma$ whereas, for moderate to large
$\gamma$, $k$ decreases: the rates turnover with the (microscopic)
friction (Figure~2). This observation can be explained with the
help of the fact that the interaction between the system (say
reactants) and the bath must transfer sufficient energy to
activate the reactants above the energy barrier leading to
products. The corresponding rate should therefore increase with
the coupling represented by friction. An increase in friction,
however, also slows down the reactants and induces a competing
mechanism that reduces the rate. Thus the topology of the
variation of $k$ with dissipation constant in the present work
also exhibits a typical signature of Kramers turnover. It is thus
important to note that the simulation of the barrier crossing
dynamics of the external noise-driven-reservoir-modulated dynamics
of the system captures the essential turnover features of the
Kramers dynamics of the closed system. In the detailed balance
principle, when instead of additive internal thermal noise (for
which FDR exists), the system encounters another multiplicative
nonthermal noise that originates due to the modulation of the bath
by an external noise, one recovers Kramers turnover nature. Thus,
the recovery of Kramers turnover for an thermodynamically open
system is the key issue of our present investigation. Figure~2
also shows that for a given value of $\gamma$, the escape rate
increases with increase in the temperature, as it should be. With
increasing temperature, the sharpness of the turnover of the
escape rate also increases.

To this end we would like to mention the works of Zhou\cite{Zhou}
and Kalmykov {\it et al.}\cite{Kalmykov}. In both of the works,
the authors have considered the standard Langevin equation with
constant and additive $\delta$-correlated white noise which
relates with the dissipation by means of FDR (and hence describe
thermodynamically closed system). In the work of Zhou\cite{Zhou},
the Langevin equation was solved numerically to study the nature
of the barrier dynamics, whereas the matrix-continued fraction
method has been exploited to examine the thermally activated
escape from a double-well potential for all values of dissipation
by Kalmykov {\it et al.}\cite{Kalmykov}. In both the works,
inevitable Kramers turnover was examined and compared with those
obtained by the Mel'nikov and Meshkov method\cite{Melnikov2}. On
the other hand, our present work deals with Kramers turnover in
the case of open system in conjunction with both additive and
multiplicative noises.

\section{Summarizing Remarks}

Many physical processes (with arbitrary complexity) influenced by
the surroundings can be modeled as a potential barrier crossing
event. Kramers showed that there is a qualitative difference in
the barrier crossing dynamics at the low and high friction limits.
Many authors have devised theoretical and computational models to
describe the Kramers turnover by extending the basic assumptions
found in the original Kramers work. The open question to be
addressed here is whether the Kramers turnover is realizable in
that class of thermodynamically open systems when the reservoir is
modulated nonlinearly by an external noise and hence is relevant
to chemical dynamics, in conjunction with other physical
processes.

This work is a continuation of our studies on the models to
describe the Kramers turnover. In Ref.~\cite{jrcbarik}, Ray
Chaudhuri {\it et al.} shown numerically that the well known
Kramers turnover phenomena is restored when the bath is {\it
linearly modulated} by an external noise. However, in the present
work the bath is being driven {\it nonlinearly} by an external
noise. In this case, in spite of having a linear system-bath
interaction, the nonthermal noise will appear multiplicatively in
the Langevin equation. The origin of this multiplicative nature
lies in the nonlinear driving of the bath itself. We have also
envisaged the Kramers turnover phenomenon for the present model.
Main results of this work are presented in Figure~2 which show the
behavior of the rate constant as functions of the friction
coefficient of the environment. From the aforesaid, we are led to
the conclusion that irrespective of the mode in which the bath
nonequilibration takes place, the turnover phenomenon will make
its appearance, and it is not only the additive noise that leads
to such an observation, but also the multiplicative noise too has
the potential to induce turnover. The observations of the present
work are valid for all types of processes in which a classical
system in contact with a thermal heat bath is driven out of
equilibrium by classical, generally time-dependent fluctuating
forces.

\section*{Acknowledgements}

The work is supported by the CSIR, India [Grant No.
01(2257)/08/EMR-II]. AS is indebted to the CSIR (Government of
India) for Senior Research Fellowship. SKB acknowledges support from Bose Institute 
through Institutional Programme VI - Development of Systems Biology.

\end{document}